# Frequency Domain Equalization for Single and Multiuser Generalized Spatial Modulation Systems in Time Dispersive Channels

Nuno Souto, *Senior Member, IEEE*, and Américo Correia, *Senior Member, IEEE*

*Abstract*—In this letter, a low-complexity iterative detector with frequency domain equalization is proposed for generalized spatial modulation (GSM) aided single carrier (SC) transmissions operating in frequency selective channels. The detector comprises three main separate tasks namely, multiple-input multiple-output (MIMO) equalization, active antenna detection and symbol wise demodulation. This approach makes the detector suitable for a broad range of MIMO configurations, which includes single-user and multiuser scenarios, as well as arbitrary signal constellations. Simulation results show that the receiver can cope with the intersymbol interference induced by severe time dispersive channels and operate in difficult underdetermined scenarios where the total number of transmitter antennas is substantially larger than the number of receiver antennas.

## I. Introduction

Generalized spatial modulation (GSM) [1]-[2] is a multiple-input multiple-output (MIMO) scheme that offers a tradeoff between the high spectral efficiency (SE) of full spatial multiplexing MIMO and the low complexity of the single radio frequency (RF) transmitter chain of spatial modulations (SMs) [3]. GSM relies on the use of multiple RF chains in order to support transmission over multiple active antenna elements (AEs). The information is mapped onto a transmit antenna combination (TAC) and on the modulated symbols, thus increasing the SE. Due to the transmission of multiple streams, GSM detection is more complex than SM. While the same symbols can be transmitted on all active AEs [1], in this letter we are concerned with the higher SE approach where a different symbol is sent on each active AE [2].

A lot of research efforts have focused on SM and GSM schemes operating in flat fading channels [1]-[6]. However, in broadband systems the channel is often severely time dispersive and leads to high intersymbol interference (ISI) levels. Although orthogonal frequency division multiplexing (OFDM) is very popular for frequency selective environments, the combination with SM sacrifices most of its benefits [7]. A better suited alternative is single carrier (SC) transmission which can potentially avoid the SM-OFDM limitations while also providing higher frequency diversity. Motivated by this, the combination of SC with SM [7]-[9] and GSM [10]-[12] recently started to attract substantial research efforts. Regarding GSM-SC, which is the focus of this letter, a low complexity detection scheme was proposed in [10]. Although it can achieve good performances in several scenarios, it was designed for zero-padded SC (ZP-SC) systems and has a complexity that grows directly with the GSM constellation size, making its application in large-scale systems difficult. In [11], several tree search algorithms for ZP-SC were proposed and evaluated. They can also achieve good performances but rely on the search over the whole GSM set making them impractical for large-scale systems. In [12] several time-domain turbo equalization detectors were proposed but were designed specifically for application to ZP-SC systems. Regarding cyclic-prefixed (CP) aided SC transmissions, most of the research has been restricted to SM only [7][9] where the special structure of CP-SC is exploited in order to implement part of the processing in the frequency domain. Against this background, the main contributions of this letter are summarized as follows:

- We develop an iterative detector for CP-aided GSM-SC systems which separates the tasks of MIMO equalization, active AE detection and symbol wise demodulation. This splitting is accomplished through the alternating direction method of the multipliers (ADMM), which we had previously applied in a simpler form in [6], within the context of single-user (SU) GSM-MIMO transmissions in flat fading channels. In this letter the detector is designed in order to cope with the more challenging multiuser (MU) scenarios and frequency selective channels.
- Through proper formulation of the SU/MU GSM-SC detection problem, the MIMO equalization is accomplished in the frequency domain which allows the complexity to remain low even in highly dispersive channels.
- Numerical results show that the receiver is suitable for arbitrary SU and MU MIMO configurations with arbitrary signal modulations, it can cope with high ISI inducing channels and operate in difficult underdetermined scenarios.

*Notation:* $(\cdot)^T$ and $(\cdot)^H$ denote the transpose and conjugate transpose of a matrix/vector, $\otimes$ symbolizes the Kronecker product, $\|\cdot\|_2$ is the 2-norm of a vector, supp$(\cdot)$ returns the set of

N. Souto and A. Correia are with the ISCTE-University Institute of Lisbon and Instituto de Telecomunicações, 1649-026 Lisboa, Portugal (e-mail: nuno.souto@lx.it.pt, americo.correia@iscte-iul.pt).

This work was funded by FCT/MEC through national funds and co-funded by FEDER – PT2020 partnership agreement under the project UID/EEA/50008/2019 and grant SFRH/BSAB/142993/2018.

indices of nonzero elements in **x** (i.e., the support of **x**), diag(·) represents a diagonal matrix, $\lfloor \cdot \rfloor$ is the floor function, $\binom{N}{k}$ denotes the number of combinations of $N$ symbols taken $k$ at a time, $\mathbf{I}_n$ is the $n \times n$ identity matrix and $I_\mathcal{D}(\mathbf{v})$ is the indicator function which returns 0 if $\mathbf{v} \in \mathcal{D}$ and $+\infty$ otherwise.

## II. SYSTEM MODEL AND PROBLEM STATEMENT

Let us consider a SC system where a base station with $N_{rx}$ receiver antennas serves $N_u$ users. Each user is equipped with $N_{tx}$ transmitter antennas with only $N_a$ active AEs at any given time. This allows a total of $N_{comb} = 2^{\lfloor \log_2 \binom{N_{tx}}{N_a} \rfloor}$ TACs available per user. Every active AE transmits a different $M$-QAM modulated symbol resulting in a total of $\lfloor \log_2 \binom{N_{tx}}{N_a} \rfloor + N_a \log_2 M$ bits mapped to each GSM symbol. A frequency selective channel with $L$ resolvable paths is assumed for each pair of transmitter-receiver antennas. We consider that the system operates with $N$-sized blocks employing a CP with length $N_{CP}$ ($N_{CP} \geq L-1$), and that the channel is time invariant during a block. The GSM signal vector transmitted by user $p$ ($p=0,\ldots,N_u-1$) during channel use $t$ ($t=-N_{CP},\ldots,N-1$) can then be expressed as

$$\mathbf{s}_t^p = \underbrace{\left[\ldots,0,s_{t,0}^p,0,\ldots,0,s_{t,N_a-1}^p,0,\ldots\right]^T}_{N_{tx}} \quad (1)$$

with $s_{t,j}^p \in \mathcal{A}$ ($j=0,\ldots,N_a-1$) and $\mathcal{A}$ denoting the $M$-sized complex valued constellation set. The received signal vector in the time domain can be written as

$$\mathbf{y}_t = \sum_{i=0}^{L-1} \mathbf{\Omega}^i \mathbf{s}_{t-i} + \mathbf{n}_t, \quad (2)$$

where $\mathbf{y}_t \in \mathbb{C}^{N_{rx} \times 1}$, $\mathbf{s}_t = \left[\mathbf{s}_t^{0\,T} \ldots \mathbf{s}_t^{N_u-1\,T}\right]^T$ and $\mathbf{n}_t \in \mathbb{C}^{N_{rx} \times 1}$ is the vector containing independent zero-mean circularly symmetric Gaussian noise samples with covariance $2\sigma^2 \mathbf{I}_{N_{rx}}$. Matrix $\mathbf{\Omega}^i \in \mathbb{C}^{N_{rx} \times N_u N_{tx}}$ contains all the channel coefficients of the $i^{th}$ tap and is defined as $\mathbf{\Omega}^i = \left[\mathbf{\Omega}^{i,0} \cdots \mathbf{\Omega}^{i,N_u-1}\right]$, where

$$\mathbf{\Omega}^{i,p} = \begin{bmatrix} h_{1,1}^{i,p} & \cdots & h_{1,N_{tx}}^{i,p} \\ \vdots & \ddots & \vdots \\ h_{N_{rx},1}^{i,p} & \cdots & h_{N_{rx},N_{tx}}^{i,p} \end{bmatrix} \quad (3)$$

and $h_{r,u}^{i,p}$ represents the complex-valued channel gain between transmit antenna $u$ of user $p$ and receive antenna $r$. Dropping the CP, we can concatenate the received vectors as $\mathbf{y} = \left[\mathbf{y}_0^T \ldots \mathbf{y}_{N-1}^T\right]^T$ and write

$$\mathbf{y} = \mathbf{\Omega}\mathbf{s} + \mathbf{n}, \quad (4)$$

where, $\mathbf{s} = \left[\mathbf{s}_0^T \ldots \mathbf{s}_{N-1}^T\right]^T$, $\mathbf{n} = \left[\mathbf{n}_0^T \ldots \mathbf{n}_{N-1}^T\right]^T$ and

$$\mathbf{\Omega} = \begin{bmatrix} \mathbf{\Omega}^0 & 0 & \cdots & \mathbf{\Omega}^{L-1} & \cdots & \mathbf{\Omega}^1 \\ \vdots & \mathbf{\Omega}^0 & \ddots & \vdots & \ddots & \vdots \\ \mathbf{\Omega}^{L-1} & \vdots & \ddots & 0 & & \mathbf{\Omega}^{L-1} \\ 0 & \mathbf{\Omega}^{L-1} & & \mathbf{\Omega}^0 & \ddots & \vdots \\ \vdots & \vdots & \ddots & \vdots & \ddots & 0 \\ 0 & 0 & \cdots & \mathbf{\Omega}^{L-1} & \cdots & \mathbf{\Omega}^0 \end{bmatrix}. \quad (5)$$

The block circulant structure of the channel matrix $\mathbf{\Omega} \in \mathbb{C}^{NN_{rx} \times NN_u N_{tx}}$ allows it to be factorized as

$$\mathbf{\Omega} = \left(\mathbf{F}^H \otimes \mathbf{I}_{N_{rx}}\right) \mathbf{H} \left(\mathbf{F} \otimes \mathbf{I}_{N_u N_{tx}}\right), \quad (6)$$

where $\mathbf{F}$ represents the unitary $N \times N$ discrete Fourier transform (DFT) matrix. $\mathbf{H}$ is a block diagonal matrix, i.e., $\mathbf{H} = \text{diag}\left(\mathbf{H}_0,\ldots,\mathbf{H}_{N-1}\right)$ with $\mathbf{H}_k = \left[\mathbf{H}_k^0 \cdots \mathbf{H}_k^{N_u-1}\right]$,

$$\mathbf{H}_k^p = \begin{bmatrix} \sum_{i=0}^{L-1} h_{1,1}^{i,p} \omega^{ki} & \cdots & \sum_{n=0}^{L-1} h_{1,N_{tx}}^{i,p} \omega^{ki} \\ \vdots & \ddots & \vdots \\ \sum_{i=0}^{L-1} h_{N_{rx},1}^{i,p} \omega^{ki} & \cdots & \sum_{i=0}^{L-1} h_{N_{rx},N_{tx}}^{i,p} \omega^{ki} \end{bmatrix}, \; k=0,\ldots,N-1, \quad (7)$$

and $\omega$ denoting a $N^{th}$ primitive root of unity. The received block can then be expressed in the frequency domain as

$$\mathbf{Y} = \left(\mathbf{F} \otimes \mathbf{I}_{N_{rx}}\right)\mathbf{y} = \mathbf{HS} + \mathbf{N}. \quad (8)$$

with $\mathbf{S} = \left(\mathbf{F} \otimes \mathbf{I}_{N_u N_{tx}}\right)\mathbf{s}$ and $\mathbf{N} = \left(\mathbf{F} \otimes \mathbf{I}_{N_{rx}}\right)\mathbf{n}$. The maximum likelihood detection (MLD) problem for the described system model can be formulated in the frequency domain as

$$\min_{\mathbf{s}} \; \|\mathbf{Y} - \mathbf{HS}\|_2^2 \quad (9)$$

$$\text{subject to} \quad \mathbf{S} = \left(\mathbf{F} \otimes \mathbf{I}_{N_u N_{tx}}\right)\mathbf{s} \quad (10)$$

$$\mathbf{s} \in \mathcal{A}_0^{NN_u N_{tx}} \quad (11)$$

$$\text{supp}\left(\mathbf{s}_t^p\right) \in \mathbb{S}, \; t=0,\ldots,N-1, \; p=0,\ldots,N_u-1, \quad (12)$$

where $\mathcal{A}_0 \stackrel{\text{def}}{=} \mathcal{A} \cup \{0\}$ and $\mathbb{S}$ denotes the set of valid TACs. Due to constraints (11) and (12), finding the exact solution requires a computational complexity that grows exponentially with the problem size making it most often impractical.

## III. FREQUENCY DOMAIN GSM DETECTOR

In this section, we apply a generalized version of ADMM [14] as a heuristic to provide good quality solutions with reduced complexity for the MLD problem. Firstly we encode constraints (11) and (12) into (9) and rewrite the problem as

$$\min_{\mathbf{S},\mathbf{z}} \; \|\mathbf{Y}-\mathbf{HS}\|_2^2 + \sum_{t=0}^{N-1}\sum_{p=0}^{N_u-1} I_{\mathbb{S}}(\mathbf{x}_t^p) + I_{\mathcal{A}_0^{NN_u N_{tx}}}(\mathbf{z}) \quad (13)$$

$$\text{subject to} \quad \mathbf{S} = \left(\mathbf{F} \otimes \mathbf{I}_{N_u N_{tx}}\right)\mathbf{z} = \left(\mathbf{F} \otimes \mathbf{I}_{N_u N_{tx}}\right)\mathbf{x}, \quad (14)$$

where $\mathbf{x} = \left[\mathbf{x}_0^T \ldots \mathbf{x}_{N-1}^T\right]^T$ with $\mathbf{x}_t = \left[\mathbf{x}_t^{0\,T} \ldots \mathbf{x}_t^{N_u-1\,T}\right]^T$. Then, the augmented Lagrangian function (ALF) can be written as

$$L_{\mathbf{P}_x,\mathbf{P}_z}(\mathbf{S},\mathbf{z},\mathbf{x},\mathbf{U},\mathbf{W}) = \|\mathbf{Y}-\mathbf{HS}\|_2^2 + \sum_{t=0}^{N-1}\sum_{p=0}^{N_u-1} I_{\mathbb{S}}(\mathbf{x}_t^p) + I_{\mathcal{A}_0^{NN_u N_{tx}}}(\mathbf{z})$$

$$+\left(\mathbf{S}-\left(\mathbf{F}\otimes\mathbf{I}_{N_uN_{tx}}\right)\mathbf{x}+\mathbf{U}\right)^H \mathbf{P}_x \left(\mathbf{S}-\left(\mathbf{F}\otimes\mathbf{I}_{N_uN_{tx}}\right)\mathbf{x}+\mathbf{U}\right)$$
$$+\left(\mathbf{S}-\left(\mathbf{F}\otimes\mathbf{I}_{N_uN_{tx}}\right)\mathbf{z}+\mathbf{W}\right)^H \mathbf{P}_z \left(\mathbf{S}-\left(\mathbf{F}\otimes\mathbf{I}_{N_uN_{tx}}\right)\mathbf{z}+\mathbf{W}\right)$$
$$-\mathbf{U}^H \mathbf{P}_x \mathbf{U} - \mathbf{W}^H \mathbf{P}_z \mathbf{W}. \quad (15)$$

where $\mathbf{U},\mathbf{W} \in \mathbb{C}^{NN_uN_{tx}\times 1}$ are the scaled dual variables while $\mathbf{P}_x = \mathrm{diag}(\rho_0^x,...,\rho_{NN_uN_{tx}-1}^x)$ and $\mathbf{P}_z = \mathrm{diag}(\rho_0^z,...,\rho_{NN_uN_{tx}-1}^z)$, with $\rho_i^x, \rho_i^z \in [0,+\infty[$, denote the penalty matrices. The gradient ascent method is then applied to the dual problem [13] resulting in the following sequence of iterative steps.

• *Step 1: Minimization of the ALF over* **S**. The frequency domain estimate at iteration $q+1$ can be obtained from $\nabla_{\mathbf{S}^H} L_{\mathbf{P}_x,\mathbf{P}_z}(\mathbf{S},\mathbf{z},\mathbf{x},\mathbf{U},\mathbf{W}) = 0$ which, exploiting the block diagonal structure of **H**, results in

$$\mathbf{S}_k^{(q+1)} = \left(\mathbf{H}_k^H \mathbf{H}_k + \mathbf{P}_{x,k} + \mathbf{P}_{z,k}\right)^{-1}\left(\mathbf{H}_k^H \mathbf{Y}_k + \mathbf{P}_{x,k}\left(\mathbf{X}_k^{(q)} - \mathbf{U}_k^{(q)}\right)\right.$$
$$\left. + \mathbf{P}_{z,k}\left(\mathbf{Z}_k^{(q)} - \mathbf{W}_k^{(q)}\right)\right), k=0,...,N-1. \quad (16)$$

$\mathbf{S}_k^{(q+1)}$, $\mathbf{Y}_k$, $\mathbf{X}_k^{(q)}$, $\mathbf{Z}_k^{(q)}$, $\mathbf{U}_k^{(q)}$, $\mathbf{W}_k^{(q)}$, $\mathbf{P}_{x,k}$ and $\mathbf{P}_{z,k}$ represent slices of $\mathbf{S}^{(q+1)}$, $\mathbf{Y}$, $\mathbf{X}^{(q)}$, $\mathbf{Z}^{(q)}$, $\mathbf{U}^{(q)}$, $\mathbf{W}^{(q)}$, $\mathbf{P}_x$ and $\mathbf{P}_z$ matching the $k^{\mathrm{th}}$ frequency. **X** and **Z** are the frequency domain representations of **x** and **z**, i.e., $\mathbf{X}=\left(\mathbf{F}\otimes\mathbf{I}_{N_uN_{tx}}\right)\mathbf{x}$ and $\mathbf{Z}=\left(\mathbf{F}\otimes\mathbf{I}_{N_uN_{tx}}\right)\mathbf{z}$.

• *Step 2: Minimization of the ALF over* **x**. This step reduces to

$$\mathbf{x}_t^{p(q+1)} = \Pi_{\mathbb{S}}\left(\mathbf{r}_t^{p(q+1)}\right), t=0,...,N-1, p=0,...,N_u-1, \quad (17)$$

where $\mathbf{r}^{(q+1)} = \left(\left(\mathbf{F}^H \otimes \mathbf{I}_{N_uN_{tx}}\right)\left(\mathbf{S}^{(q+1)}+\mathbf{U}^{(q)}\right)\right)$ and $\Pi_{\mathbb{S}}(\cdot)$ denotes the projection onto $\mathbb{S}$.

• *Step 3: Minimization of the ALF over* **z**. In this case we get

$$\mathbf{z}^{(q+1)} = \Pi_{\mathcal{A}_0^{NN_uN_{tx}}}\left(\left(\mathbf{F}^H\otimes\mathbf{I}_{N_uN_{tx}}\right)\left(\mathbf{S}^{(q+1)}+\mathbf{W}^{(q)}\right)\right), \quad (18)$$

where $\Pi_{\mathcal{A}_0^{NN_uN_{tx}}}$ denotes the projection onto $\mathcal{A}_0^{NN_uN_{tx}}$ which can be implemented as a simple rounding of each component to the closest element in $\mathcal{A}_0$.

• *Step 4: Dual variable update*. The update of the dual variables is accomplished through

$$\mathbf{U}^{(q+1)} = \mathbf{U}^{(q)} + \mathbf{S}^{(q+1)} - \mathbf{X}^{(q+1)}, \quad (19)$$
$$\mathbf{W}^{(q+1)} = \mathbf{W}^{(q)} + \mathbf{S}^{(q+1)} - \mathbf{Z}^{(q+1)}. \quad (20)$$

Algorithm 1 summarizes all the required steps, with $\hat{\mathbf{s}}$ denoting the final estimate and $Q$ the maximum number of iterations. In lines 11-14, $I$ is the support of $\mathbf{x}^{(q+1)}$, $\bar{I}$ is the respective complement (i.e., $\bar{I}=\{1,...,NN_uN_{tx}\}\setminus I$), and $\hat{\mathbf{s}}_I$ ($\hat{\mathbf{s}}_I^{\mathrm{candidate}}$) is the reduced $N_aN_uN\times 1$ vector containing the nonzero elements of $\hat{\mathbf{s}}$ ($\hat{\mathbf{s}}^{\mathrm{candidate}}$) given by the support $I$. For initialization of the algorithm we can perform a random selection of a vector **s** with elements constrained within the constellation limits, followed by the projection over $\mathbb{S}$ and $\mathcal{A}_0^{NN_uN_{tx}}$ in order to obtain $\mathbf{x}^0$ and $\mathbf{z}^0$. $\mathbf{U}^0$ and $\mathbf{W}^0$ can be set as 0. To improve the chance of finding the optimal solution the algorithm can be run multiple times with different initializations [14]. The penalty coefficients, $\rho_i^x$ and $\rho_i^z$, are used as tuning parameters for achieving the best performance for a specific problem setting. Regarding the implementation of the algorithm, the multiplications by $\left(\mathbf{F}\otimes\mathbf{I}_{N_uN_{tx}}\right)$ and $\left(\mathbf{F}^H\otimes\mathbf{I}_{N_uN_{tx}}\right)$ can be efficiently performed through $N_uN_{tx}$ fast Fourier transforms (FFTs) which results in a complexity order of $O\left(NN_u^3N_{tx}^3 + NQN_u^2N_{tx}^2 + QN_uN_{tx}N\log_2 N\right)$.

---

**Algorithm 1:** Proposed Frequency Domain ADMM based Detector for GSM-MIMO (GSM-FD-ADMM)

1: **Input:** $\mathbf{Y}$, $\mathbf{H}$, $\mathbf{U}^0$, $\mathbf{W}^0$, $\mathbf{x}^0$, $\mathbf{z}^0$, $\mathbf{P}_x$, $\mathbf{P}_z$, $Q$
2: $f_{best} = \infty$.
3: $\mathbf{X}^0 \leftarrow \left(\mathbf{F}\otimes\mathbf{I}_{N_uN_{tx}}\right)\mathbf{x}^0$, $\mathbf{Z}^0 \leftarrow \left(\mathbf{F}\otimes\mathbf{I}_{N_uN_{tx}}\right)\mathbf{z}^0$.
4: **for** $q=0,1,...Q-1$ **do**
5:   Compute $\mathbf{S}_k^{(q+1)}$ using (16) for all frequencies $k=0,1,...,N-1$.
6:   Obtain $\mathbf{x}_t^{p(q+1)}$ for all symbol positions and users with projection (17).
7:   Obtain $\mathbf{z}^{(q+1)}$ with projection (18).
8:   $I \leftarrow \mathrm{supp}\left(\mathbf{x}^{(q+1)}\right)$
9:   $\hat{\mathbf{s}}_{\bar{I}}^{\mathrm{candidate}} \leftarrow 0$, $\hat{\mathbf{s}}_I^{\mathrm{candidate}} \leftarrow \Pi_{\mathcal{A}^{N_aN_uN}}\left(\mathbf{s}_I^{(q+1)}\right)$.
10:   **If** $f\left(\hat{\mathbf{s}}^{\mathrm{candidate}}\right) < f_{best}$ **then**
11:     $\hat{\mathbf{s}}_{\bar{I}} \leftarrow 0$, $\hat{\mathbf{s}}_I \leftarrow \hat{\mathbf{s}}_I^{\mathrm{candidate}}$.
12:     $f_{best} = f\left(\hat{\mathbf{s}}^{\mathrm{candidate}}\right)$.
13:   **end if**
14:   $\mathbf{X}^{(q+1)} \leftarrow \left(\mathbf{F}\otimes\mathbf{I}_{N_uN_{tx}}\right)\mathbf{x}^{(q+1)}$,
    $\mathbf{Z}^{(q+1)} \leftarrow \left(\mathbf{F}\otimes\mathbf{I}_{N_uN_{tx}}\right)\mathbf{z}^{(q+1)}$
15:   $\mathbf{U}^{(q+1)} = \mathbf{U}^{(q)} + \mathbf{S}^{(q+1)} - \mathbf{X}^{(q+1)}$.
16:   $\mathbf{W}^{(q+1)} = \mathbf{W}^{(q)} + \mathbf{S}^{(q+1)} - \mathbf{Z}^{(q+1)}$
17: **end for**
18: **Output:** $\hat{\mathbf{s}}$.

---

## IV. NUMERICAL RESULTS

In this section, we evaluate the performance of the proposed detector using Monte Carlo simulations. An uncoded MU SC system with $N=128$, a block duration of 67μs and a CP with 16.7μs was considered. The adopted channel model was the Extended Typical Urban model (ETU) [15] (similar conclusions could be drawn for other severely time-dispersive channels). All the channel coefficients were independently drawn according to a zero-mean complex Gaussian distribution. Randomly selected modulated symbols were transmitted on the active AEs with $E\left[|s_i|^2\right]=1$.

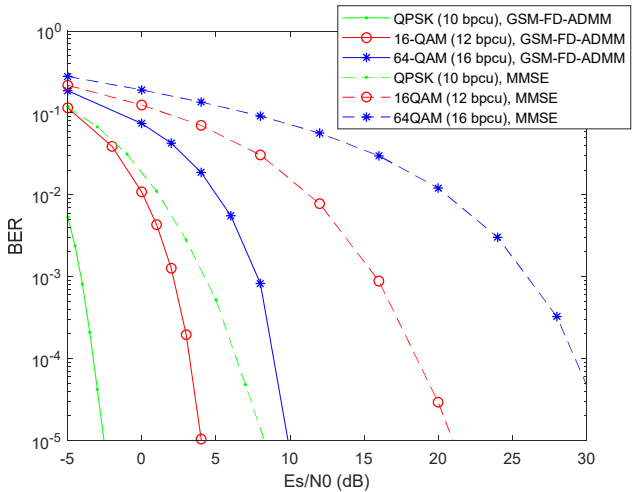

Fig. 1. BER performance of a multiuser SC-GSM-MIMO system with $N_u$=12, $N_{tx}$=7, $N_a$=2, $N_{rx}$=84 and different modulations.

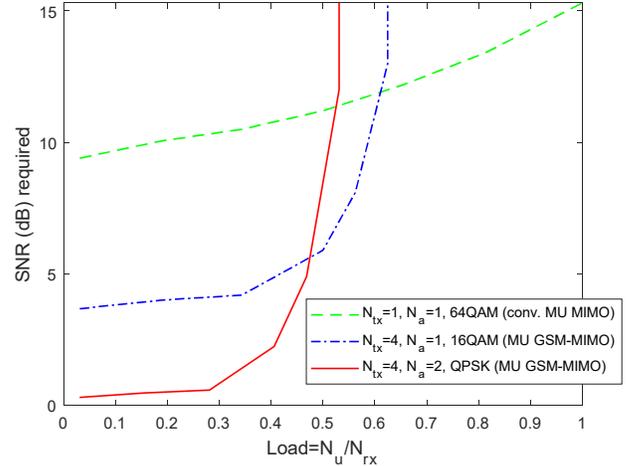

Fig. 2. SNR required for achieving a target BER of $10^{-4}$ versus loading factor ($N_u/N_{rx}$) considering $N_{rx}$=32 and 6 bpcu per user.

Fig. 1 plots the bit error rate (BER) as a function of the signal to noise ratio (SNR) per receive antenna, for a MU scenario with $N_u$=12, $N_{tx}$=7, $N_a$=2 and $N_{rx}$=84 and different modulations resulting in different SE in bits per channel use (bpcu). The proposed detector was applied with 5 random initializations and $Q$=30 iterations. The penalty parameters values were $\rho_i^x = \rho_i^z = 60$. As a reference, MMSE curves are also included. We can observe that the proposed receiver is able to effectively cope with the ISI induced by the channel and at the same time detect the GSM symbols, providing very significant gains over the MMSE (10 dB for QPSK and 20 dB for 64-QAM at a BER of $10^{-4}$).

Fig. 2 illustrates the impact of changing the loading factor (defined as $N_u/N_{rx}$) on the SNR required to achieve a target BER of $10^{-4}$ when $N_{rx}$=32. Three different configurations with the same SE of 6 bpcu per user are considered. The proposed receiver is employed for all setups, including the case $N_{tx}=N_a=1$ (conventional MU MIMO). For low loads, the use of GSM has a clear performance advantage over the conventional MU system. For loads above 0.25, the GSM systems becomes underdetermined ($N_u N_{tx} > N_{rx}$) and, even though the proposed receiver is still able to perform well, the SNR degradation becomes sharper until the point where the BER of $10^{-4}$ becomes unreachable. In this high-load region, the conventional MU setup becomes a better performing solution.

## V. Conclusions

This letter presented a novel iterative detector for SU and MU SC-GSM transmissions in frequency-selective channels which accomplishes reduced complexity implementation through frequency domain equalization. Numerical simulations show that the proposed receiver can effectively cope with the ISI induced by severe time dispersive channels and operate in difficult underdetermined scenarios. The inherent splitting-based design of the algorithm allows it to easily deal with GSM based transmissions, which can be more attractive in low load scenarios, and switch to conventional MU detection whenever the load becomes high.